\documentstyle[emulateapj,epsfig,apjfonts,epsf]{article}
 
\begin{document}

\newcommand{\maxima}{{\sc Maxima}} \newcommand{\boom}{{\sc Boomerang}}
\def\fun#1#2{\lower3.6pt\vbox{\baselineskip0pt\lineskip.9pt
    \ialign{$\mathsurround=0pt#1\hfil##\hfil$\crcr#2\crcr\sim\crcr}}}
\def\lta{\mathrel{\mathpalette\fun <}}    % \lta = less than approx.
\def\gta{\mathrel{\mathpalette\fun >}}    % \gta = greater than approx. 
                             
\title{Implications for quintessence models from \maxima-1 and
  \boom-98}

\author {A. Balbi\altaffilmark{1}, C. Baccigalupi\altaffilmark{2}, S.
  Matarrese\altaffilmark{3,4}, F. Perrotta\altaffilmark{2}, N.
  Vittorio\altaffilmark{1}}

\altaffiltext{1}{Dipartimento di Fisica, Universit\`a Tor Vergata,
  Roma, I-00133, Italy} \altaffiltext{2}{SISSA/ISAS, Via Beirut 4,
  34014 Trieste, Italy} \altaffiltext{3}{Dipartimento di Fisica
  `Galileo Galilei', Universit\'a di Padova, and INFN, Sezione di
  Padova, Via Marzolo 8, 35131 Padova, Italy}
\altaffiltext{4}{Max-Planck-Institut f\"ur Astrophysik,
  Karl-Schwarzschild-Strasse 1, D-85748 Garching, Germany}

\begin{abstract}
  Prompted by the recent \maxima-1 and \boom-98 measurements of the
  cosmic microwave background (CMB) anisotropy power spectrum, and
  motivated by the results from the observation of high-redshift Type
  Ia supernovae, we investigate CMB anisotropies in quintessence
  models in order to characterize the nature of the dark energy today.
  We perform a Bayesian likelihood analysis, using the \maxima-1 and
  \boom-98 published bandpowers, in combination with COBE/DMR, to
  explore the space of quintessence parameters: the quintessence
  energy density $\Omega_\phi$ and equation of state $w_\phi$.  We
  restrict our analysis to flat, scale-invariant, inflationary
  adiabatic models. We find that this simple class of inflationary
  models, with a quintessence component $\Omega_\phi\lta 0.7$, $-1\leq
  w_\phi\lta-0.5$, is in good agreement with the data. Within the
  assumptions of our analysis, pure quintessence models seem to be
  slightly favored, although the simple cosmological constant scenario
  is consistent with the data.
\end{abstract}

\keywords{Cosmic microwave background --- Cosmology: observations ---
  Cosmology: theory --- equation of state}

\section{Introduction}

Recent results (Perlmutter et al. 1999; Riess et al. 1998) from the
observation of high-redshift Type Ia supernovae (SNe Ia) have driven a
renewed interest towards cosmological scenarios where the bulk of the
energy density is made up of some unknown negative-pressure component
(named ``dark energy'' in the recent literature).  Among the
possibilities that have been proposed to explain the nature of this
component is the existence of a non-zero cosmological constant
(constant vacuum energy).  A more general way to describe this
component is by its equation of state $p=w\rho$, with $w<0$. This
allows for a whole set of candidates with different values of $w$.
Among them is a dynamical form of time varying, spatially
inhomogeneous dark energy component, generally provided by a slowly
rolling scalar field, referred to as ``quintessence'' (Caldwell et al.
1998). The most attractive feature of quintessence models is their
capability to solve the so-called `cosmic coincidence' problem, namely
the fact that matter and vacuum energy contribute by comparable
amounts to the energy density today.  Successful quintessence models
should in fact allow for the existence of attractor (Ratra \& Peebles
1988) or `tracker' (Zlatev et al. 1999) solutions, which make the
present-day behavior of the scalar field nearly independent of its
initial conditions (e.g. Wang et al. 2000, for a thorough review).

It has been noticed (see, e.g., Perlmutter, Turner \& White 1999) that
it may be difficult for SNe Ia data alone to discriminate among
different candidates for the dark energy.  Complementing these data
with those coming from other observations may help assessing the
nature of the dark energy component.  Study of the cosmic microwave
background (CMB) anisotropy has long been recognized as a powerful
tool to investigate cosmological models. The recent outstanding
achievements of the \maxima-1 and \boom-98 experiments (Hanany et al.
2000; de Bernardis et al. 2000) have finally moved us closer to the
goal of a long awaited high precision measurement of the CMB angular
power spectrum.  The CMB power spectrum measurements by \maxima-1 and
\boom-98 are in remarkable agreement, and have already been used to
set constraints on the values of a suite of cosmological parameters
within the class of inflationary adiabatic models (Jaffe et al. 2000;
Balbi et al. 2000; Lange et al. 2000).  These results, which strongly
support a universe with a density very close to critical, made the
case for the existence of dark energy even stronger. In fact, large
scale structure observations indicate that matter with strong
clustering properties can only make up to $\sim 30\%$ of the critical
density, leaving room for a $\sim 70\%$ contribution from dark energy.
It is then timely to use these new CMB datasets to explore in more
detail the nature of the dark energy.
A comparison between quintessence models and recent CMB data has been
performed by Brax et al. (2000); they investigated the dependence of
CMB power spectrum (in particular the location and height of acoustic
features) on the main cosmological parameters as well as on different
quintessence potentials. The predicted CMB anisotropy is then compared
to \boom-98 and \maxima-1, finding general agreement.  A likelihood
analysis of the \boom-98 results for theories involving couplings of
the quintessence field with other cosmological components has been
carried on by Amendola (2000), finding again broad compatibility with
the data.
In this {\it Letter} we focus on the quintessence scenario, and use
the CMB power spectrum from \maxima-1 and \boom-98 to set limits to
the parameters of this model. These are the present-day closure energy
density of the quintessence component, $\Omega_\phi$, and the quantity
$w_\phi\equiv p_\phi/\rho_\phi$, characterizing its equation of state.
We have purposely restricted our analysis to the simplest class of
inflationary theories, namely flat models with scale-invariant
adiabatic scalar perturbations.

This {\it Letter} is organized as follows. In Section 2 we give a
quick review of the key features of the quintessence scenario, with a
particular emphasis to the theoretical predictions for the CMB angular
power spectrum. In Section 3 we use the \maxima-1 and \boom-98 power
spectrum measurements to constrain the parameters of the model.
Finally, we discuss our results in Section 4.

\section{CMB Anisotropy in quintessence models}

The quintessence component is usually described as an ultra-light (its
Compton wavelength being of the same order or larger than the Hubble
radius) self-interacting scalar field $\phi$, with negative pressure.
It is invoked to provide the accelerated expansion of the Universe
today in alternative to the ordinary cosmological constant.  In this
scenario, the vacuum energy is stored in the potential energy $V(\phi
)$ of the $\phi$ field, evolving through the classical Klein-Gordon
equation. A suitable period of slow-rolling in the scalar field
dynamics allows to reproduce a behavior similar to a cosmological
constant, which in turn requires that the potential should admit a
rather flat region.  Several candidates have been proposed in recent
years in the particle physics context (e.g. Masiero, Pietroni \&
Rosati 2000, and refs. therein). In this Letter we focus on
quintessence models with inverse power-law potentials, $V\propto
\phi^{-\alpha}$ ($\alpha >0$), as originally suggested by Ratra \&
Peebles (1988).

The large interest in quintessence models is due to their capability
to avoid the severe fine-tuning on the smallness of the vacuum energy
compared with matter and radiation in the early universe: a large
class of field trajectories, named ``tracking solutions'' (Zlatev et
al. 1999), are able to set the $\phi$ energy at the observed value
today starting from a very wide set of initial conditions; among
others, Ratra-Peebles-type potentials do satisfy this requirement
(Liddle \& Scherrer 1999).

For a given class of potentials, quintessence models can be
parametrized by two quantities: the scalar field closure energy
density today, $\Omega_{\phi}$, and its equation of state at the
present time, $w_\phi=[p_\phi/\rho_\phi]_{t_0}$, where $p_\phi$ and
$\rho_\phi$ are the field pressure and energy density, respectively.
Notice that $w_\phi=-1$ for a pure cosmological constant, but can be
larger for scalar field component. In particular, in the tracking
regime $\rho_\phi \propto (1+z)^{3\alpha /(\alpha +2)}$ at low
redshifts, thus reducing to the ordinary cosmological constant for
$\alpha\rightarrow 0$.  Moreover, there is a one-to-one relation
between $w_\phi$ and the potential exponent $\alpha$:
$w_{\phi}=-2/(\alpha +2)$: the larger $w_{\phi}$, the greater the
inverse potential exponent $\alpha$.  However, one must be aware that,
although useful for a qualitative understanding, the above relations
are only approximately satisfied at present, since the quintessence
field is dominating and the tracking regime is being abandoned.

Even though the $\phi$ field is too relativistic to cluster on
sub-horizon scales, its dynamics induces relevant imprints on
cosmological scales. The scales beyond the horizon at decoupling are
depicted on the low-$\ell$ tail of the CMB anisotropy angular power
spectrum; the quintessence dynamics acts on these scales through the
Integrated Sachs-Wolfe effect (ISW; e.g. Hu, Sugiyama \& Silk 1997).
Also, the $\phi$ component participates in the overall expansion of
the Universe, causing a geometric effect that translates into a
projection of each physical scale onto different angles in the sky.
Moreover, since its energy density was largely subdominant at
decoupling, the form of the acoustic peaks region of the CMB spectrum
at $\ell\ge 200$ is not altered.

For a given value of the vacuum energy density today, the effects of a
quintessence component, compared to the cosmological constant, can be
qualitatively understood by considering the dark energy equation of
state.  When moving away from $w_{\phi}=-1$, the first effect that is
important is the projection one: the size of the last scattering
surface gets reduced because in the past the quintessence energy
density was larger than that of a cosmological constant: this
determines a shift of all the CMB features to larger scales, or
smaller multipoles.  In addition, the more $w_{\phi}$ deviates from
$-1$, the earlier is the time when $\phi$ starts to dominate the
cosmic expansion; through the ISW effect, the CMB photons are
sensitive to the dynamics of the gravitational potential wells and
hills along their trajectories, which in turn decay much more rapidly,
reflecting the dominance of the vacuum energy. This enhances the level
of CMB anisotropies on the low $\ell$'s tail of the spectrum,
where one should consider the limitations introduced by cosmic
variance, when dealing with the ISW power. However, for the tracking
solutions analysed here, the variation in the power spectrum
introduced by the ISW for $-0.6 \lta w_\phi\lta -0.5$ is larger than the
uncertainty due to cosmic variance.

CMB anisotropy angular power spectra in the considered class of
quintessence models have been obtained by suitably modifying CMBFAST
(Seljak \& Zaldarriaga 1996) to include $\phi$ perturbations in the
synchronous gauge, as detailed in Perrotta \& Baccigalupi (1999), and
numerically evolving the background quantities, accounting for the
tracker scalar field trajectories, as discussed in Baccigalupi,
Matarrese \& Perrotta (2000), where results were presented for some
specific values of the quintessence parameters.

\vskip 0.5cm \vbox{\hskip -0.3cm\epsfxsize=9cm\epsfbox{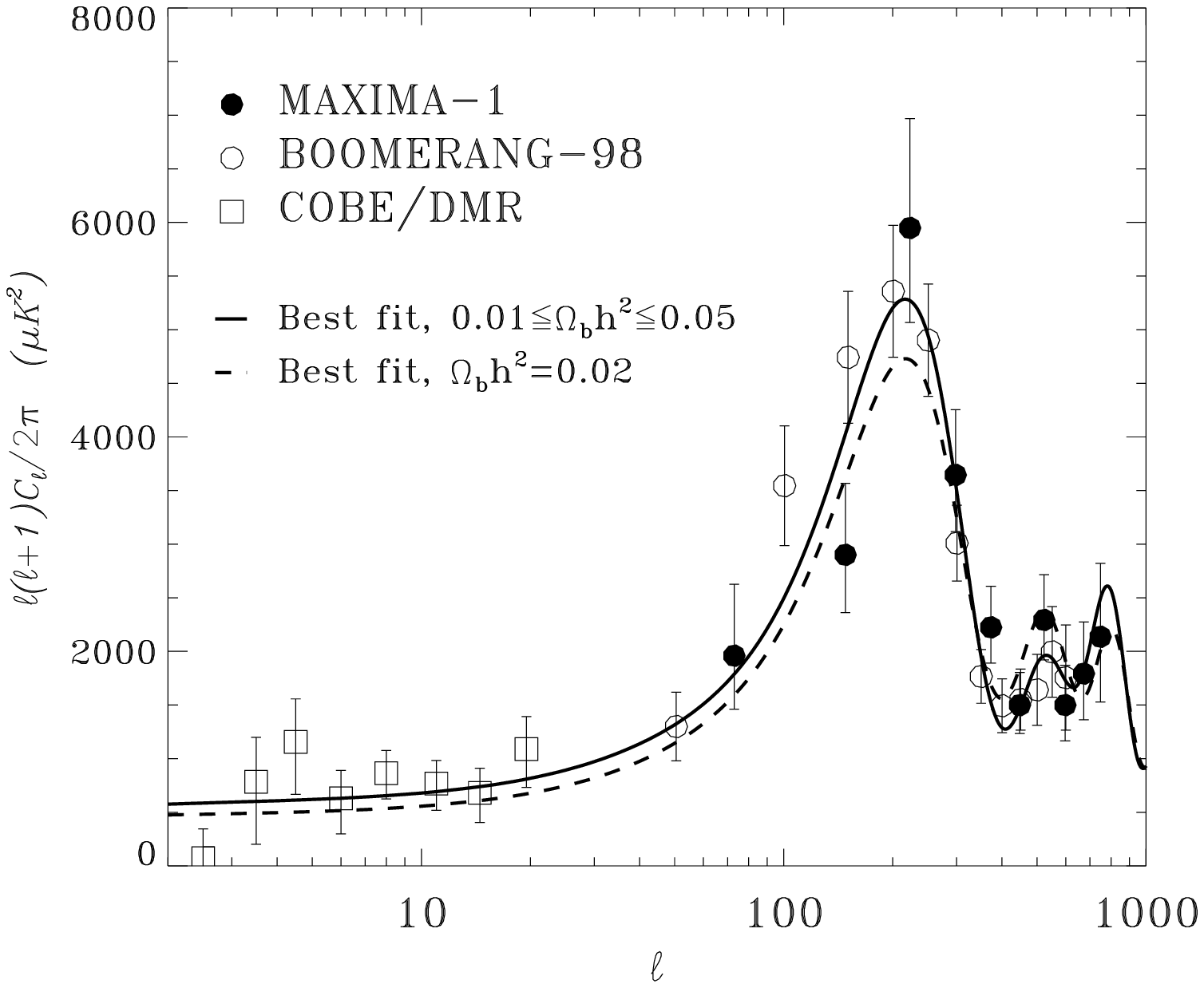}} \vskip
0.5cm { \small F{\scriptsize IG}.~1.--- Data points used in the
  analysis. The full circles are the bandpowers from \maxima-1, the
  open circles are from \boom-98.
  These data have been re-scaled within their 1$\sigma$ calibration
  uncertainties.  The open boxes are the COBE/DMR data.
  The curves are the theoretical predictions from quintessence models
  which best fit the combined data when we allow the physical baryon
  density to vary in the range $0.01\leq\Omega_b h^2\leq 0.05$
  (continuous line, corresponding to $\Omega_\phi=0.5$, $w_\phi=-0.75$
  and $\Omega_b h^2= 0.03$) or we fix it at the big bang
  nucleosynthesis value $\Omega_b h^2=0.02$ (dashed line,
  corresponding to $\Omega_\phi=0.7$ and $w_\phi=-0.75$).}  \vskip
0.5cm

\section{Comparison with the data} 

In this section we compare the predictions of quintessence models to
the results of the first flight of the \maxima~experiment (\maxima-1;
Hanany et al.~2000) and of the long duration Antarctic flight of
\boom~(\boom-98; de Bernardis et al.~2000).  To rule out models in the
2-dimensional space of quintessence parameters ($\Omega_\phi$,
$w_\phi$), we adopted a Bayesian approach, estimating the probability
distribution of the parameters by evaluating the likelihood of the
data, ${\cal L}\propto \exp{(-\chi^2/2)}$.  The data we used are the
\maxima-1 spectrum, estimated in 10 bins covering the range
$36\leq\ell\leq 785$, and the \boom-98 spectrum, estimated in 12 bins,
over the range $26\leq\ell\leq 625$. Within each bin, the power
spectrum is parameterized by a flat bandpower: ${\cal
  C}_b\equiv\ell(\ell+1)C_\ell$.  We used the published bandpowers and
neglected the effect of bin-bin correlations (which should be small,
see Hanany et al. 2000, de Bernardis et al. 2000), and the
offset-lognormal corrections to the likelihood, since these have not
been made public yet.  Following Jaffe et al. (2000), we combined the
\maxima-1 and \boom-98 adjusting the calibration of each dataset by a
factor $0.98\pm 0.08$ for \maxima-1 and $1.14\pm 0.10$ for \boom-98.
We extended the low $\ell$'s coverage of the dataset by including in
our analysis the COBE/DMR power spectrum estimated over 8 bins in the
range $2\leq\ell\leq 22$ by Tegmark \& Hamilton (1997).  The band
powers used in the analysis are shown in Figure~1.

Since the aim of this work is to focus specifically on the
quintessence parameters space, we did not embark in a
multi-dimensional fit such as those already performed in Lange et al.
(2000), Balbi et al.  (2000), Jaffe et al. (2000). We intentionally
decided to restrict our analysis to the most basic set of inflationary
adiabatic models, namely those with a flat geometry
($\Omega\equiv\Omega_\phi+\Omega_m=1$) and a scale-invariant spectrum
of primordial fluctuations ($n_s=1$).  Furthermore this class of
inflationary models is precisely the one which seems to be favored by
the current multi-parameter CMB analyses (see also, e.g., Kinney,
Melchiorri \& Riotto, 2000).  We treated the overall normalization of
the power spectrum as a free parameter, and for each model we adjusted
it to the value which gave the best fit to our data.  We fixed the
value of the Hubble constant at $H_0=65$~km~s$^{-1}$~Mpc$^{-1}$, which
is consistent with most of the observations (Freedman 1999).  We
assumed that reionization occurred late enough to have a negligible
effect on the power spectrum (corresponding to an optical depth
$\tau=0$). Finally, we did not include the effect of massive neutrinos
(which would be very small anyway) and of tensor modes (gravitational
waves).

We only made an exception to our strategy of using very tight priors
in the case of the physical baryon density $\Omega_b h^2$ (where $h$
is the Hubble constant in units of 100~km~s$^{-1}$~Mpc$^{-1}$), since
the value $\Omega_b h^2=0.019\pm 0.002$ deduced from big bang
nucleosynthesis (BBN) considerations using the measured deuterium
abundances (Burles et al.  1998) seems to conflict with the value
$\Omega_b h^2=0.032\pm 0.005$ (Jaffe et al. 2000) emerging from the
already mentioned analyses of the CMB datasets (see also Esposito et
al. 2000). Then, in order to test the importance of this discrepancy
in our analysis, we allowed the physical baryon density to vary in the
range $0.01\leq\Omega_b h^2\leq 0.05$, which encompasses a more
conservative range consistent with high and low deuterium abundances
measurements. In this case our results were obtained by marginalizing
(integrating) the likelihood over $\Omega_b h^2$.

Our results are shown in Figure~2 where we plot the 68\% and 95\%
confidence levels in the $\Omega_\phi$--$w_\phi$ plane.  When we let
$\Omega_b h^2$ vary, we are able to find quintessence models which are
an excellent fit to our data. We point out that most of the weight to
the likelihood is provided by models with $\Omega_b h^2\sim 0.03$.
Indeed, marginalizing over $\Omega_b h^2$ or fixing it at $\Omega_b
h^2= 0.03$ give almost exactly the same result. The best fit is a
model with $\Omega_\phi=0.5$, $w_\phi= - 0.75$, $\Omega_b h^2= 0.03$,
having $\chi^2$/DOF=26/27 (where DOF indicates the degrees of freedom,
resulting from 30 data points and 3 parameters).  If we fix the
physical baryon density to the BBN value $\Omega_b h^2=0.02$, the
bounds we obtain in the $\Omega_\phi$--$w_\phi$ plane are
qualitatively similar to those of, e.g., Perlmutter, Turner \& White
(1999), Efstathiou (1999), Wang et al. (2000).  However, we stress the
fact that none of the models with $\Omega_b h^2=0.02$ provides a good
fit to the data. In fact, the best fit, with $\Omega_\phi=0.7$ and
$w_\phi=-0.75$, has $\chi^2$/DOF=38/27.  Qualitatively, we found that
increasing $\Omega_b h^2$ has the effect of moving the best fit to
lower values of $\Omega_\phi$, in a roughly linear fashion, while
$w_\phi$ remains roughly constant at $w_\phi\sim-0.75$.  The best fit
models with $\Omega_b h^2=0.03$ and $\Omega_b h^2= 0.02$ are compared
to the data in Figure~1.

\vskip 0.5 cm \vbox{\hskip -0.3cm\epsfxsize=9cm\epsfbox{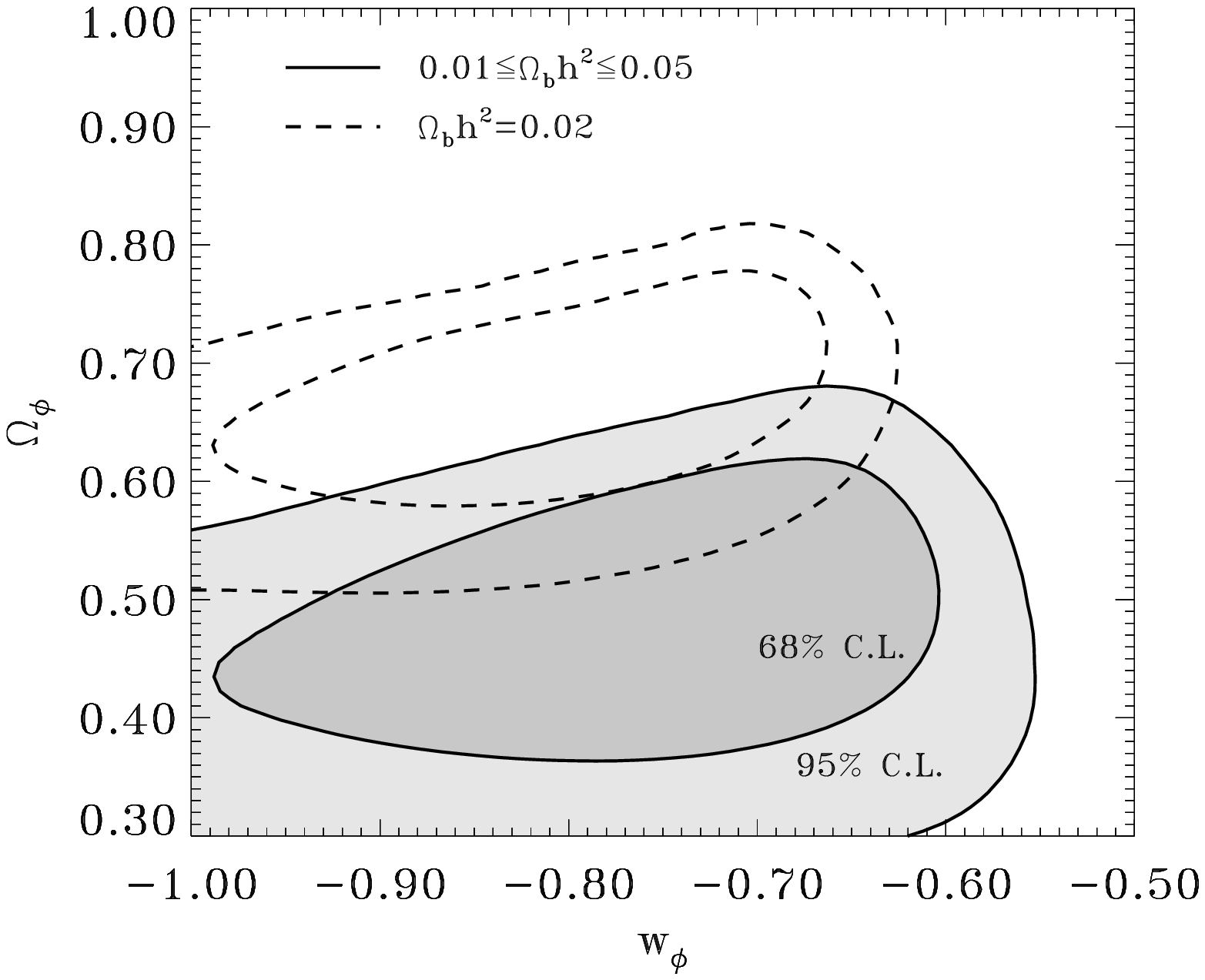}}
\vskip 0.5cm { \small F{\scriptsize IG}.~2.--- Likelihood contours at
  $68\%$ and $95\%$ confidence level in the $\Omega_\phi$---$w_\phi$
  plane.  The shaded regions are obtained marginalizing over $\Omega_b
  h^2$ in the range $0.01\leq\Omega_b h^2\leq 0.05$.  For comparison,
  the dashed contours show the bounds obtained imposing the big bang
  nucleosynthesis value $\Omega_b h^2=0.02$.}  \vskip 0.5 cm
\vbox{\hskip -0.3cm\epsfxsize=8.5cm\epsfbox{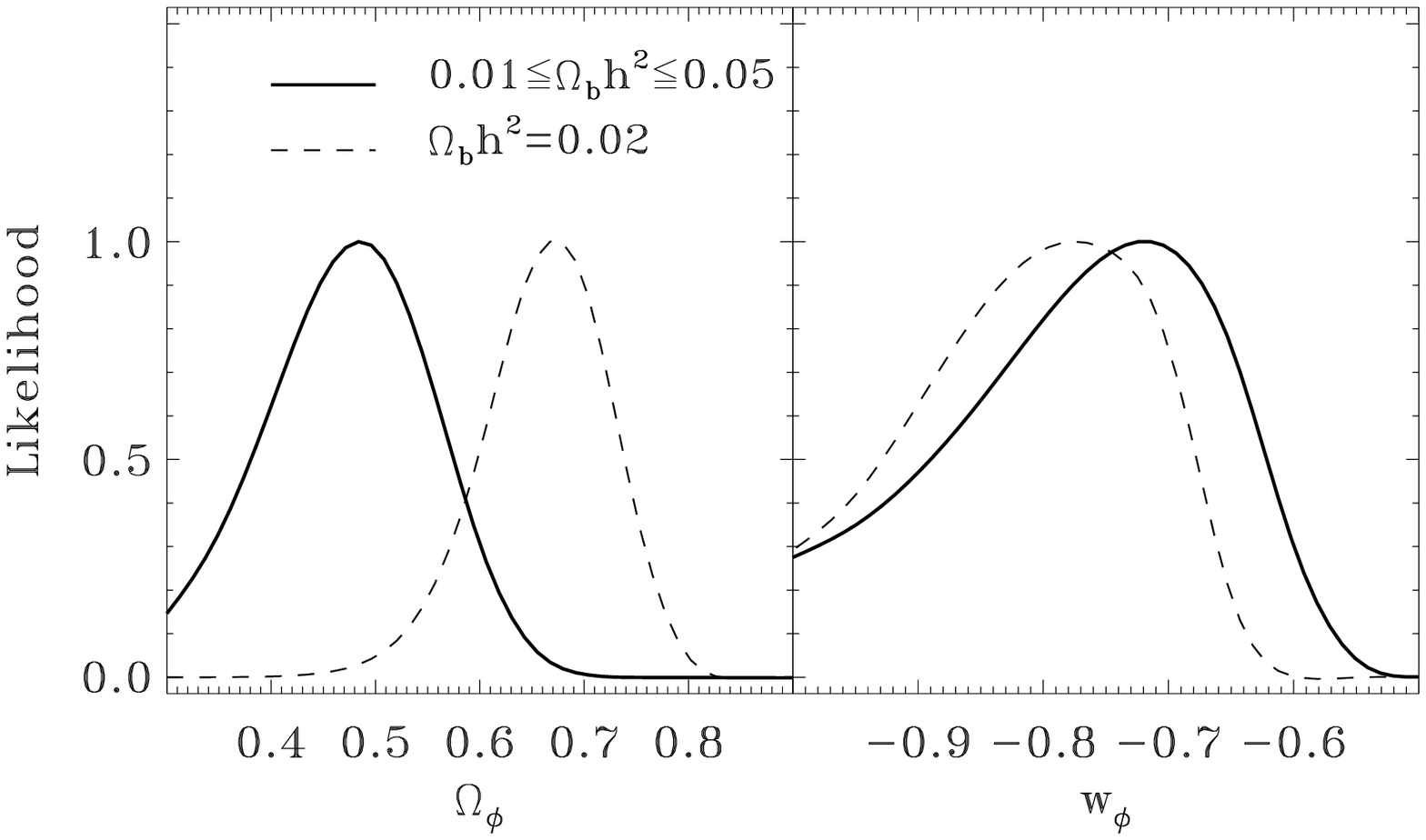}} { \small
  F{\scriptsize IG}.~3.--- Marginalized likelihood, normalized to the
  peak value, for the individual quintessence parameters $\Omega_\phi$
  and $w_\phi$.  In both panels, the solid line is obtained
  marginalizing over $\Omega_b h^2$ in the range $0.01\leq\Omega_b
  h^2\leq 0.05$, while the dashed line is obtained imposing the big
  bang nucleosynthesis value $\Omega_b h^2=0.02$. Note, however, that
  the absolute likelihood obtained for $\Omega_b h^2=0.02$ would be
  more than two orders of magnitude lower than that obtained
  marginalizing over $\Omega_b h^2$ (see text).}  \vskip 0.5 cm

In Figure~3 we show the projected likelihood functions for the
individual quintessence parameters, obtained by marginalization of the
2-dimensional likelihood with respect to each parameter.  A robust
indication, quite independent of the value of $\Omega_b h^2$, is that
although pure cosmological constant models ($w_\phi=-1$) are
compatible with the CMB data used here, pure quintessence models
($w_\phi>-1$) seem to be slightly favored.  This appears to agree with
the qualitative discussion of Brax, Martin \& Riazuelo (2000). The
peak location in models with $w_\phi>-1$ is shifted to the left with
respect to models with $w_\phi=-1$, resulting in better agreement with
the data. In fact, this mechanism allows one to find quintessence
models with $\Omega=1$ which have the same peaks locations of
non-quintessence models with a $\Omega$ slightly larger than unity.
If we impose $w_\phi=-1$, we get a best fit with $\Omega_\phi=0.45$
and $\Omega_b h^2=0.03$ with $\chi^2/$DOF=28/27 ($\Omega_\phi=0.65$ if
we impose $\Omega_b h^2=0.02$, with $\chi^2$/DOF=40/27).  It is also
interesting to note that these data seem to disfavor high values of
$\Omega_\phi$, unless we impose the BBN constraint on $\Omega_b h^2$.
However, we should keep in mind that this result is influenced by the
assumptions we made on the value of the other parameters.  Relaxing
the priors would result in higher upper bounds on $\Omega_\phi$,
similar to those found in Lange et al. (2000) or Jaffe et al (2000).

\section{Discussion and Conclusions}

We used the recent CMB power spectrum measurements from \maxima-1 and
\boom-98, combined with COBE/DMR, to investigate the quintessence
models in the context of flat, scale-invariant, inflationary adiabatic
models.  The \maxima-1 and \boom-98 data already provide interesting
insights on the equation of state of the universe. The simplest class
of inflationary models, with quintessence, together with economic
assumptions on the other parameters, provide an excellent fit to these
CMB data. In fact, quintessence models seem to be slightly favored
over simple cosmological constant models.

In this {\it Letter} we restricted our attention to the simplest class
of quintessence potentials, having the Ratra-Peebles $V \propto
\phi^{-\alpha}$ form. Within this class of models one can already
conclude that shallower potentials ($\alpha \lta 2$) are favored by
current CMB anisotropy data. Also, we did not analyze here the case
when the scalar field providing the dark energy today is non-minimally
coupled to the gravitational sector of the theory, as it happens in
the `extended quintessence' scenario (Perrotta, Baccigalupi \&
Matarrese 2000; Baccigalupi et al. 2000).  We did not compare
explicitly the constraints on the equation of state coming from
high-redshift SNe Ia observations (Perlmutter, Turner \& White 1999;
Garnavich et al. 1998) with the results of our analysis, because of
our choice of considering a range of values for $\Omega_b h^2$ not
limited to the standard BBN value.  Clearly, high precision SNe Ia
distance measurements such as those expected to come from the proposed
SNAP\footnote{http://snap.lbl.gov} satellite will provide powerful
constraints on quintessence models complementary to those of the
MAP\footnote{http://map.gsfc.nasa.gov} and
Planck\footnote{http://astro.estec.esa.nl/Planck/} CMB missions.
Similar complementary information will come from measurements of the
comoving density of galaxy clusters (Haiman, Mohr \& Holder 2000).
Thus, one will be able to put severe constraints not only on the
nature of the dark energy, but also on the structure of the theory of
gravity.

\acknowledgments We thank Matthias Bartelmann for thoughtful comments.

\end{document}